\let\myover=\over      
\documentstyle[prd,aps,graphics,epsfig,floats]{revtex}
\def\const{\mbox{const}}

\def\l{\left(}
\def\r{\right)}

\newcommand{\be}{\begin{equation}}
\newcommand{\ee}{\end{equation}}

\newcommand{\bg}{\begin{gather}}
\newcommand{\eg}{\end{gather}}

\def\half{\frac{1}{2}}

\begin{document}
\let\over=\myover  
\def\half{{1 \over 2}} 
\twocolumn[\hsize\textwidth\columnwidth\hsize\csname
@twocolumnfalse\endcsname

\tighten

\title{\vskip 0.5cm
Axion-Like Particles as Ultra High Energy Cosmic Rays?}

\author{D.S.~Gorbunov$^{(1)}$, G.G.~Raffelt$^{(2)}$, 
D.V.~Semikoz$^{(1,2)}$}
\address{$^{(1)}${\it Institute for Nuclear Research of the 
Academy of Sciences of Russia, Moscow 117312, Russia}}
\address{$^{(2)}${\it Max-Planck-Institut f\"ur Physik
(Werner-Heisenberg-Institut), F\"ohringer Ring 6,
80805 M\"unchen, Germany}}

\maketitle

\begin{abstract}
  If Ultra High Energy Cosmic Rays (UHECRs) with $E>4 \times
  10^{19}$~eV originate from BL Lacertae at cosmological distances as
  suggested by recent studies, the absence of the GZK cutoff can not
  be reconciled with Standard-Model particle properties.  Axions would
  escape the GZK cutoff, but even the coherent conversion and
  back-conversion between photons and axions in large-scale magnetic
  fields is not enough to produce the required flux.  However, one may
  construct models of other novel (pseudo)scalar neutral particles
  with properties that would allow for sufficient rates of particle
  production in the source and shower production in the atmosphere to
  explain the observations. As an explicit example for such particles
  we consider SUSY models with light sgoldstinos.
\end{abstract}
\vskip2pc]

\section{Introduction}

Ultra High Energy Cosmic Rays (UHECRs) with energies above the 
Greisen-Zatsepin-Kuzmin (GZK)
cutoff~\cite{GZK} were detected in all relevant
experiments~\cite{AGASA,fy,hp,sugar,yk}, suggesting that these
particles can not originate at cosmological distances.  On the other
hand, there are no apparent nearby sources in their arrival direction.
Therefore, something fundamental appears to be missing in our
understanding of the sources, nature, or propagation of UHECRs.

The small-scale clustering of UHECR events suggests that the sources
are point-like on cosmological scales~\cite{Tinyakov:2001ic}.  Several
astrophysical sources were suggested based on the coincidence of the
arrival directions of some of the highest-energy events with certain
astrophysical objects~\cite{elbert}. For example, a correlation
between compact radio quasars and UHECRs was suggested
in~\cite{Farrar:1998we,Farrar:1999fw,virmani}, although other authors
found them to be insignificant~\cite{Hoffman:1999ev,sigl}.  Recently,
a statistically significant correlation, at the level of chance
coincidence below $10^{-5}$, was found with the most powerful BL
Lacertae, i.e.~quasars with beams pointed in our
direction~\cite{Tinyakov:2001b}.  The identified sources are at $z >
0.1$, far exceeding the GZK distance of $R_{\rm GZK}\approx 50$~Mpc,
so that the primary UHE particles can not be protons.  The photon
attenuation length for energies around $10^{20}$~eV is of order the
GZK cutoff distance, primarily due to the extragalactic radio
backgrounds.  While the limiting magnitude of the radio backgrounds
necessary to absorb UHE photons can be determined only by numerical
propagation codes~\cite{kks1999}, one can even now conclude that
UHECRs with energies around $10^{20}$~eV are very unlikely to be
photons.

The only Standard-Model particles which can reach our Galaxy without
significant loss of energy are neutrinos.  Two different scenarios
involving UHE neutrinos have been proposed.  In the first, neutrinos
produce nucleons and photons via resonant $Z$-production with relic
neutrinos clustered within about 50~Mpc from the Earth, giving rise to
angular correlations with high-redshift sources~\cite{weiler}.
However, for the interaction rates to be sufficiently high, this
scenario requires enormous neutrino fluxes and an extreme clustering
of relic neutrinos with masses in the eV range~\cite{zburst}.  The
second neutrino scenario invokes increased high-energy
neutrino-nucleon cross sections.  This could be caused by the exchange
of Kaluza-Klein graviton modes in the context of extra
dimensions~\cite{extradim} or by an exponential increase of the number
of degrees of freedom in the context of string theory~\cite{string}.

Another possibility to avoid the GZK cutoff is a small violation of
Lorentz-invariance, a hypothesis which can not be tested in
terrestrial experiments~\cite{Lorentz,Sigl98}.

The GZK cutoff can be avoided also if the UHECRs consist of certain
new particles.  One possibility is a new stable massive hadron with a
mass around 2--3~GeV~\cite{Farrar:1996rg}, shifting the GZK bound to
higher energies $E> 10^{21}$~eV into a range where no UHECR event has
yet been found.  However, it now appears that these exotic hadrons are
excluded by laboratory experiments~\cite{gluino}.

Therefore, if the UHECRs indeed originate from point sources at
cosmological distances one is running dangerously short of plausible
explanations for how this radiation can reach us. This perhaps
desperate situation motivates us to consider other options for new
particles which can traverse the universe unimpeded at high energies.
Specifically, we consider the possibility of axion-like particles,
i.e.\ electrically neutral (pseudo)scalar particles $X$ with a
relatively small mass $M_X<10$~MeV.

Such particles must fulfill several requirements to be candidates for
UHECRs.  They must live long enough to reach us from a cosmological
distance.  They must not lose too much energy in interactions with the
CMBR and other background radiations or in extragalactic magnetic
fields.  They must interact sufficiently strongly in or near our
Galaxy or in the Earth's atmosphere to produce the observed UHE
events. Finally, their interactions must allow for the production of a
significant flux at the source.
 
We will first consider proper axions and find that they seem to be
excluded as UHECRs.  We then turn to more general particles and study
their necessary properties to fulfill the above requirements. As an
explicit example we study light sgoldstinos.

\section{Proper Axions}

Proper axions arise from the Peccei-Quinn mechanism to solve the
strong CP problem. As such their properties are governed by one main
parameter, the Peccei-Quinn scale or axion decay constant $f_a$;
astrophysical limits imply $f_a \agt 10^{10}$~GeV.  Axions mix with
neutral pions so that their mass and interaction strength are roughly
those of $\pi^0$, reduced by $f_\pi/f_a$ with $f_\pi\approx 93~{\rm
  MeV}$ the pion decay constant.  It is easy to see that axions live
long enough and interact weakly enough with the CMBR to traverse
cosmological distances unimpeded. By the same token, their interaction
strength is far too weak to imagine their efficient production at the
source or their efficient detection in the Earth's atmosphere.

It is less obvious, however, if they could not be produced in
sufficient numbers by their coherent conversion $\gamma\to a$ in
large-scale magnetic fields in the source region, and then re-appear
as photons in the galaxy by the inverse process.  Put another way, one
might imagine the UHECRs to be photons which traverse the universe in
the guise of axions.

The conversion between axions and photons in a large-scale magnetic
field is essentially a particle oscillation phenomenon~\cite{book}.
The diagonal elements of the mixing matrix involve $m_a^2$ and the
square of the ``photon effective mass'' within the given medium, the
off-diagonal element, which induces the mixing, is $2g_{a\gamma}B E$
where $B$ is the transverse magnetic field and $E$ the particle
energy.  

The oscillation length $\ell_{\rm osc}$ corresponds to the momentum
difference between axions and photons of the given energy, in our case
$E\approx 10^{20}~{\rm eV}$. Noting that the effective photon mass is
much smaller than $m_a$, the momentum difference is governed by the
axion mass alone so that $\ell_{\rm osc}= 4\pi E/m_a^2= 8.1~{\rm
  kpc}\,(E/10^{20}{\rm eV})\,({\rm meV}/m_a)^2$.  On the other hand,
the coherence length $\ell_B$ of the galactic magnetic field is
probably less than 1~kpc.  A significant conversion rate requires
$\ell_{\rm osc}\alt \ell_B$, i.e.~$m_a$ larger than a few meV, not in
contradiction with current limits.

The effective mixing angle between axions and photons in a magnetic
field is given by $\frac{1}{2}\tan(2\theta)=g_{a\gamma} B E/m_a^2
=0.2\,(g_{a\gamma}/10^{-10}{\rm GeV}^{-1})\, (E/10^{20}{\rm
  eV})\,({\rm meV}/m_a)^2$.  The astrophysical limit on the
axion-photon coupling is $g_{a\gamma}<10^{-10}~{\rm GeV}^{-1}$ so that
for $m_a$ not much smaller than 1~meV the mixing angle becomes large.
Put another way, for $g_{a\gamma}$ near its limit and $m_a$ around
1~meV the transition rate in the galaxy is not ridiculously small.

The numbers are much worse for proper axions where $g_{a\gamma}$ and
$m_a$ are related by $g_{a\gamma}
\approx \alpha/(2\pi f_a)$ and $m_a\approx m_\pi f_\pi/f_a$.  In this
case the mixing angle becomes $\theta\approx2\times10^{-4}
(E/10^{20}{\rm eV})\,({\rm meV}/m_a)\ll 1$.  Allowing the axion mass
to be very small, the mixing angle could become reasonably large.  On
the other hand, the oscillation length then becomes very much larger
than $\ell_B$.  The transition probability is $P(a\to
\gamma)=(g_{a\gamma} B \ell_B/2)^2 \approx3\times10^{-8}\,(B/\mu{\rm
  G})^2\,(\ell_B/{\rm kpc})^2\, (10^{10}{\rm GeV}/f_a)^2\ll 1$.  A
similar estimate applies to the source region where the magnetic field
could be stronger, but the correlation length would be smaller.
Therefore, the combined probability to produce axions at the source
and to convert them into photons in the Galactic magnetic field is
tiny, perhaps as small as $P \sim 10^{-16}$. Therefore, if one adjusts
the proton flux from the sources to the observed flux below the GZK
cutoff, then no significant axion flux will be produced above the
cutoff.  Therefore, proper axions do not work for this scenario.

Even if one dials $g_{a\gamma}$ and $m_a$ independently, the numbers
look discouraging as one would need a huge rate of UHE photon
production in the source to compensate for small transition rates both
in the source and galaxy and one would need parameter values near
their exclusion limits.

\section{Generic Axion-Like Particles} 
\label{gen-ax-like}

Since proper axions are apparently not able to explain the UHECR
phenomenon, we next turn to a more exotic new scalar $X$; a similar
analysis for pseudoscalars is straightforward.  The new particle is
assumed to couple to gluons and photons via nonrenormalizable
interactions of the form~\footnote{The axion-photon coupling of the
  previous section was based on the normalization ${\cal
    L}_{a\gamma}=(g_{a\gamma}/4) a F\widetilde F =g_{a\gamma} a {\bf
    E}\cdot{\bf B}$.}
\begin{equation}
{\cal L}=g_g X G_{\mu\nu}^aG^{\mu\nu}_a\;,~~~~
{\cal L}=g_\gamma X F_{\mu\nu}F^{\mu\nu}\;.
\label{Xgg}
\end{equation}
Only these two interactions will be important, so we assume that the
coupling to other Standard-Model particles are suppressed because,
say, they proceed through loops or are proportional to small Yukawa
constants.

If $M_X<2m_\pi=270$~MeV, the dominant decay mode is into two
photons:
\begin{equation}
\Gamma(X\to\gamma\gamma)={g_\gamma^2M_X^3\over4\pi}\; ,
\label{X-to-2photons}
\end{equation}
because the direct coupling to electrons is suppressed by
assumption.
If this light particle has the energy $E_X$ it propagates through the
Universe without decay if
\begin{equation}
R_{\rm Universe}\lesssim L_{\rm decay} = {E_X\over \Gamma_XM_X},
\end{equation}
where $\Gamma_X$ is essentially identical with the two-photon decay
rate Eq.~(\ref{X-to-2photons}).  Therefore, we need to require
\begin{equation}
g_\gamma\lesssim 1.6\times10^{-11}~{\rm GeV}^{-1}\,
\sqrt{E_X\over 10^{20}~{\rm eV}}\l{10~{\rm
MeV}\over M_X}\r^2
\label{X-lifetime-gamma}
\end{equation}
if these particles are supposed to reach us from cosmological 
distances.

Propagating through the Universe, the light scalar $X$ may also
disappear by interactions with the CMBR.  For $E_X \approx
10^{20}$~eV, the CM energy is $E_{{\rm cm}} \approx (2 E_X
\omega_0)^{1/2} \approx 350$~MeV, where $\omega_0 \approx 6\times
10^{-4}$~eV is the average energy of relic photons.  Pairs of light
charged particles $A^\pm$ are produced with the cross section
$\sigma(X\gamma\to A^+A^-)=\alpha g_\gamma^2/16$.  With a relic photon
number density of about $400$~cm$^{-3}$ the requirement $R_{X\gamma\to
  A^+A^-}> R_{\rm Universe}$ gives $g_\gamma<1$~GeV$^{-1}$.  Similar
estimates apply to other possible processes like $X\gamma_{{\rm CMB}}
\rightarrow \gamma \pi^0$.  Therefore, the tiny photon coupling
required by Eq.~(\ref{X-lifetime-gamma}) guarantees the absence of a
GZK cutoff for the $X$ particles.

Both the production of $X$ particles at the source and their
interaction in the atmosphere require rather large cross sections,
comparable to strong ones.  For $X$ particles with the characteristic
energy scale $g_g^{-1}$ this is possible only if the CM energy in the
system is close to this scale, but not significantly higher so that
the effective interactions~(\ref{Xgg}) are still meaningful.  Typical
CM energies of UHECR interactions with nucleons are $E_{\rm cm}
\approx 100$--300~TeV.  We can estimate the interaction cross section
with nucleons at such energies as
\begin{equation}
\sigma_X = \sigma_s ~\frac{\alpha_X}{\alpha_s}\;.
\label{crosssection}
\end{equation}
The suppression factor
\begin{equation}
 \frac{\alpha_X}{\alpha_s} = \frac{(E_{\rm cm}g_g)^2}{4\pi\alpha_s}  
\label{suppress}
\end{equation}
should not be very small. 

We next turn to the $X$ mean free path (mfp) $\ell_X$ in the Earth's
atmosphere. Since our particle exhibits strong interactions we
estimate $\ell_X$ by analogy with the proton mfp $\ell_p$ as
$\ell_X=\ell_p\, (\alpha_s/\alpha_g)$. To initiate an atmospheric
shower, $X$ should have a relatively small mfp.  Assuming
$\ell_X<10\,\ell_p$ and using Eq.~(\ref{suppress}) and $\alpha_s=0.1$
we estimate
\begin{equation}
g_g>1.1\times 10^{-6}~{\rm GeV}^{-1}\,
\sqrt{{10^{20}~{\rm eV}\over E_X}}\;.
\label{g-X-atmosphere}
\end{equation}
The inequalities~(\ref{X-lifetime-gamma}) and~(\ref{g-X-atmosphere})
determine the $g_g$ range suitable for explaining the UHECRs above the
GZK cutoff.

How are the $X$-particles produced at an astrophysical source like a
quasar?  If our estimate for the cross section
Eq.~(\ref{crosssection}) is valid, UHE $X$ particles will be
efficiently produced in the high-energy tail of the proton spectra by
proton-proton collisions while their production at low energies will
be negligible.  Therefore, we can expect that the proton flux from the
source at low energies will continue with the same slope at high
energies due to the $X$ component. Only part of the initial proton
energy will be transfered to the $X$ particles; probably they will be
produced on the peak of the gluon distribution function with $E
\approx 0.1 E_p$.  However, once produced they will escape more easily
from the source compared with protons precisely because their cross
section is smaller.

\begin{figure}[ht]
\centering\leavevmode\epsfxsize=3.4in\epsfbox{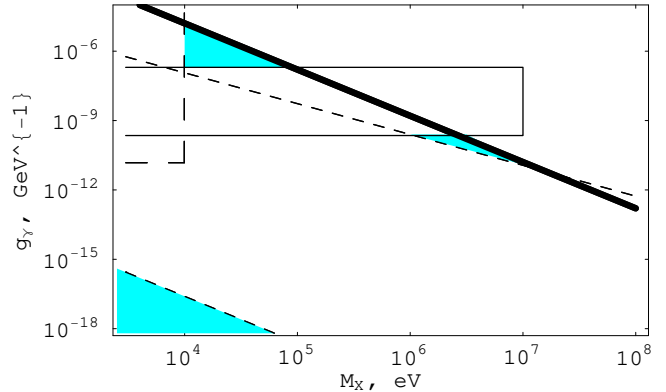}
\caption{The allowed region for the parameters $(M_X,g_\gamma)$ are
shaded in grey.  The region traced by the long-dashed line is ruled
out by the helium-burning life-time of horizontal-branch stars~\protect\cite{book}. The
region surrounded by a thin solid line is ruled out by SN~1987A. The
region confined between short-dashed lines is ruled out by the photon
background and the CMBR~~\protect\cite{distortion}. Below the thick solid line the
inequality~(\ref{X-lifetime-gamma}) is valid.}
\label{general-fig}
\end{figure}

Many bounds on axion-like particles arise from cosmology, astrophysics
and laboratory measurements~\cite{pdg,distortion}.  Still, there remain regions
in parameter space where $X$ particles can explain UHECRs without
contradicting these limits. In Fig.~\ref{general-fig} we present the
experimentally allowed regions in the space ($g_\gamma$,$M_X$) where
the inequality~(\ref{X-lifetime-gamma}) is satisfied.  In each
concrete model one can evaluate the effective coupling constant
$g_\gamma$ which has to belong to the allowed regions shown in
Fig.~\ref{general-fig}. Since generally the interaction with gluons
leads at higher order to an effective interaction with photons, the
inequality~(\ref{g-X-atmosphere}) may shrink the allowed regions in
Fig.~\ref{general-fig} in concrete models.

  From the general case one can see that constraints on the $X$
particle interactions favor a strong coupling to gluons and a tiny one
to photons.  Hence the first extreme example is a light scalar $X$
which interacts at tree level only with gluons according to
Eq.~(\ref{Xgg}); a similar analysis applies to a light
pseudoscalar. The interaction with all other SM particles arises at
higher order.  In particular, because the gluonic operator creates
mesonic fields, the interaction $X\gamma\gamma$ emerges with a
coupling constant respecting the hierarchy
$g_\gamma/g_g\sim\alpha/(4\pi)\sim10^{-3}$.  In view of this
relationship the inequality~(\ref{g-X-atmosphere}) allows only the
region of parameter space which corresponds to the upper shaded region
in Fig.~\ref{general-fig}. Unfortunately, this allowed region
corresponds to a fairly small $g_g^{-1}\sim0.1$--5~TeV.  Therefore,
our nonrenormalizable model for $X$-baryon scattering in the
atmosphere becomes invalid because it should proceed at 100~TeV in the
CM frame.

This example shows that the lowest region in Fig.~\ref{general-fig} is
unphysical, because the condition ~(\ref{g-X-atmosphere}) requires the
hierarchy $g_\gamma/g_g\sim 10^{-10}$, which is impossible due to loop
contributions. The $M_X \sim$~MeV region in Fig.~\ref{general-fig}
can still exist in models with a hierarchy between photon and gluon
couplings, but this requires a two order of magnitude fine-tuning for
the ratio $g_\gamma/g_g$ down to values of order~$10^{-5}$.

The other possibility is that the couplings to photons and to gluons
are of the same order. In this case only the upper region in
Fig.~\ref{general-fig} is interesting because the gluon coupling
should not be too small from Eq.~(\ref{g-X-atmosphere}).  We now turn
to an explicit example for a model which does not need any fine tuning
of the couplings $g_\gamma$ and $g_g$.

\section{Light Sgoldstinos}

As an example of a realistic model for $X$ particles we consider the
supersymmetric extension of the SM with a light scalar and/or
pseudoscalar sgoldstino, the superpartner of the goldstino.  The
sgoldstino couplings are $g_g=M_3/(2\sqrt{2}F)$ and
$g_\gamma=M_{\gamma\gamma}/(2\sqrt{2}F)$, where $F$ is a parameter of
supersymmetry breaking and
$M_{\gamma\gamma}=M_1\cos^2\theta_W+M_2\sin^2\theta_W$ with $M_i$ the
corresponding gaugino masses.  Therefore, the sgoldstino coupling to
photons is suppressed relative to gluons only by the ``hierarchy among
gauginos.'' Therefore, this is an example for a model where $X$
couples to photons with a similar strength as to gluons.  For
$M_3=5M_{\gamma\gamma}=500$~GeV we obtain
\begin{equation}
\sqrt{F}\gtrsim 1.5\times10^6~{\rm GeV}
\left({10^{20}~{\rm eV}\over E_X}\right)^{1/4}
{M_X\over 10~{\rm MeV}}
\label{S-lifetime-gamma}
\end{equation}
instead of Eq.~(\ref{X-lifetime-gamma}) and 
\begin{equation}
\sqrt{F}\lesssim
1.3\times 10^4~{\rm GeV}
\left({E_X\over10^{20}~{\rm eV}}\right)^{1/4}
\label{g-S-atmosphere}
\end{equation}
instead of Eq.~(\ref{g-X-atmosphere}).

\begin{figure}[ht]
\centering\leavevmode\epsfxsize=3.4in\epsfbox{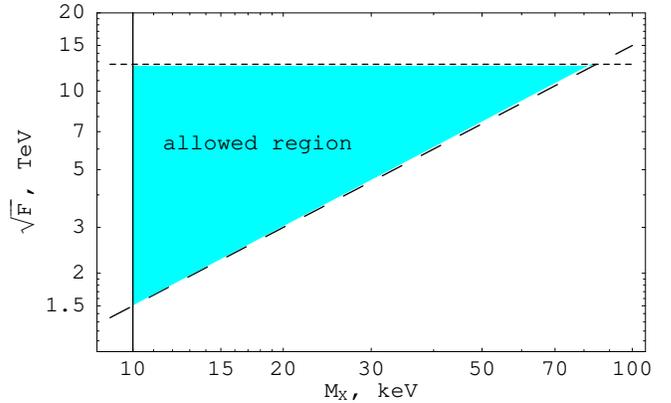}
\caption{Allowed region for the parameters
  $(M_X,\sqrt{F})$. The short-dashed line corresponds to the
  limit~(\ref{g-X-atmosphere}), the long-dashed line
  to~(\ref{X-lifetime-gamma}).  Sgoldstinos with masses less than
  10~keV (vertical solid line) are ruled out by the helium-burning
  life-time of horizontal-branch stars.}
\label{fig:ex2}
\end{figure}

A variety of experimental limits on models with light sgoldstinos has
been derived in~\cite{Gorbunov:2000th}.  In Fig.~\ref{fig:ex2} we
present the region of parameter space where sgoldstinos may act as
UHECRs and are not excluded by other limits. This region corresponds
to the upper region in Fig.~\ref{general-fig}.

If $E_X=10^{21}$~eV or more, the allowed regions are larger,
though no event of such energies has been observed.  If
$g_s=\const/\Lambda$ where $\Lambda$ is the scale of new physics, then
at $\const\sim1$ we have $\Lambda=10^2$--$10^3$~TeV.  With
$E_X=10^{11}$~GeV we have $E_{\rm cm}=300$~TeV for interactions with
protons.  Certainly $\Lambda$ should exceed this value if we want to
use the nonrenormalizable interactions~(\ref{Xgg}). For sgoldstinos we
have $M_{\rm soft}\sim \const~F/\Lambda$ and $\Lambda$ should be
larger than $E_{\rm cm}=300$~TeV. Note that $F$ is a parameter of
supersymmetry breaking and $\Lambda$ is something like the scale of
mediation of supersymmetry breaking which generally differs from
$\sqrt{F}$ but should exceed $\sqrt{F}$ if $\const$ is of order~1.

\section{Conclusions}

We have suggested new (pseudo)scalar particles as Ultra High Energy
Cosmic Rays beyond the GZK cutoff.  Our analysis was particularly
motivated by recent results suggesting that the sources of UHECRs are
cosmologically point-like~\cite{Tinyakov:2001ic} and that at least
some of the sources appear to be BL~Lacertae~\cite{Tinyakov:2001b} at
cosmological distances.

We have calculated the required range of parameters characterizing
these particles if we postulate that they should be produced in
high-redshift sources, propagate through the Universe without decay or
energy loss, and interact in the Earth's atmosphere strongly enough to
produce extended air showers at energies beyond the GZK cutoff.  The
self-consistency of our analysis requires that the energy scale for
new physics, which for SUSY models is the scale of mediation of
supersymmetry breaking, should be close to the UHECR center-of-mass
energy with nucleons of $E_{\rm cm}=300$~TeV.

As a specific example we studied light
sgoldstinos.  We considered restrictions on the parameters of the 
model which come from laboratory experiments and observational data.
We obtained the required region in parameter space of the model
which obeys all existing limits.

We note that our allowed region in Fig.~2 suggests that the
supersymmetry breaking scale $\sqrt{F} \sim 1$--10~TeV. Hence our
light sgoldstino model can be tested in searches for rare decays of
$J/\psi$ and $\Upsilon$ and in reactor experiments (for details see
Ref.~\cite{Gorbunov:2000th}).  This low scale of supersymmetry
breaking may be also tested at new generation accelerators like
Tevatron and LHC.  Also, sgoldstino contributions to FCNC and lepton
flavor violation are strong enough to probe the supersymmetry breaking
scale up to $\sqrt{F}\sim10^4$~TeV~\cite{Gorbunov:2000th} if
off-diagonal entries in squark (slepton) mass matrices are close to
the current limits in the MSSM. Thus our light-sgoldstino scenario for
UHECRs allows only small flavor violation in the scalar sector of
superpartners.
 
Light (pseudo)scalars emerge not only in the context of supersymmetry,
but also, for instance, in string theory and 
models with extra dimensions. Probably, such scalars also can serve as UHECRs if their
effective coupling with photons obeys the limits presented in
section~\ref{gen-ax-like}.

Interpreting the UHECRs as new (pseudo)scalars is, of course,
extremely speculative. However, we think it is noteworthy that such an
interpretation is at all possible and self-consistent without
violating existing limits.

\section*{Acknowledgments}

We thank B.~Ermolaev, V.~Kuzmin, A.~Neronov, V.~Rubakov, L.~Stodolsky,
I.~Tkachev and V.~Zakharov for helpful discussions and comments.  The
work of GR and DS was supported, in part, by the Deutsche
Forschungsgemeinschaft under grant No.\ SFB-375.  The work of DG was
supported in part by RFBR grant 99-01-18410, by the CPG and SSLSS
grant 00-15-96626, by CRDF grant (award RP1-2103) and by SSF grant
7SUPJ062239.

\end{document}